\documentclass[article,showpacs,twocolumn,aps]{revtex4}
\usepackage[dvips,final]{graphicx}
\begin{document}
\draft

\title{Controlled coupling of selected single quantum dots with high-Q microdisk microcavities}
\author{V. Zwiller\footnote{Now at: Institute of Quantum Electronics and Photonics, EPFL, CH-1015 Lausanne, Switzerland\\ e-mail:Valery.Zwiller@epfl.ch}, S. F\"alth\footnote{Now at: Institute of Quantum Electronics, ETH H\"onggeberg HPT G16, CH-8093 Z\"urich, Switzerland}, M. T. Bj\"ork, W. Seifert, L. Samuelson}
\address{Solid State Physics, Lund University, Box 118, SE-22100 Lund, Sweden}

\author{G. Bj\"ork}
\address{Department of Microelectronics and Information Technology, Royal Institute of Technology (KTH), Electrum 229, SE-16440 Kista, Sweden }

\date{\today }

\
\begin{abstract}
We have fabricated and studied the photoluminescence from microdisks containing single, selected self-assembled quantum dots. Using two electron beam lithography exposures and a two-step selective wet etching process, the dots were positioned at the microdisks edges with sub-micron precision. The selection and positioning of quantum dots enables an optimum coupling of the dot emission to the microdisks whispering modes.
\end{abstract}



\maketitle


Microdisks offer a very efficient optical confinement at their periphery, yielding high quality factors \cite{McCall,Slusher}. Q-values as high as 17000 have been reported for microdisks with InAs quantum dots \cite{MichlerDisk} and values as high as $10^8$ have been reached with SiO$_2$ microdisks \cite{VahalaNature}. This results in sharp emission lines and low lasing thresholds \cite{CaoMDLasing,MichlerMDLasing}. Quantum dots are especially well suited for microdisks as they introduce less absorption than quantum wells. Coupling a single quantum dot to the whispering modes of a microdisk would enable an optimum coupling of a single emitter to the maximum quality factor value that can be attained. Single quantum dot based devices offer new functionalities such as single photon emission \cite{MichlerScience,Becher,Kiraz,Zwiller,ZwillerRed}, a single dot is an exciting tool to study cavity quantum electrodynamics effects, as has been done with single atoms in cavities \cite{Rempe}. The controlled coupling of a single emitter to a high quality factor cavity also enables a careful study of lifetime modifications. The inherent random positioning of self-assembled quantum dots makes it difficult to position dots in a cavity. All lifetime modification studies reported so far have been done with randomly positioned dots in cavities \cite{Gerard,Solomon,GayralPurcell,KirazPurcell}. A precise control  of the dot position enables an optimal study of the emitter-cavity coupling \cite{VuckovicPC}, we achieve this goal by positioning the cavity respective to the dots. Other techniques that have been used to position microstructures in relation to selected QDs \cite{Falth} was considered unsuitable for producing high-quality microdisks.

The sample was grown by metal organic vapor phase epitaxy on a GaAs wafer. A GaAs buffer layer was first deposited followed by a 100 nm thick GaInP layer serving as an etch stop during the pillar etching, and 600 nm of GaAs was then deposited, in order to set the height of the pedestal. This value was chosen to avoid optical coupling with the substrate while still providing an efficient thermal contact between the microdisk and the substrate. Finally, 90 nm of GaInP were deposited with a low density of InP dots in the center of the layer. 

A coordinate system was fabricated on the sample using electron beam lithography. Gold marks were produced in regular arrays by thermal evaporation of gold followed by liftoff. The sample was then mapped with a micro-photoluminescence setup with sub-micron resolution. The photoluminescence from single dots and white light reflection from the metal markers was imaged simultaneously. These images gave the position of quantum dots relative to the metal markers, an image is shown in Fig. 1a.

A second electron beam lithography exposure was made to define the microdisks. The gold marks were easily imaged through the resist with the electron microscope and were used as alignment markers to produce disks containing the selected dots in their periphery, where the whispering modes are most intense. Circular SiO masks with 5 micron diameter were obtained following liftoff. The structures were produced by wet etching to obtain high Q values \cite{Gayral}. A non selective etchant was used to etch through the GaInP and into the GaAs before the SiO mask was removed. The pedestal was then formed by selective etching. Details of the processing have been published elsewhere \cite{OurLaser}. A complete microdisk structure containing a single quantum dot is shown in Fig. 1c. Fig. 1 a and b were taken on the same location, during mapping and after completing the processing. The dots are numbered and comparing the two images clearly shows that all the dots have been successfully placed at the microdisks edges.

\begin{figure}[h]
\center{\includegraphics[width=1\columnwidth]{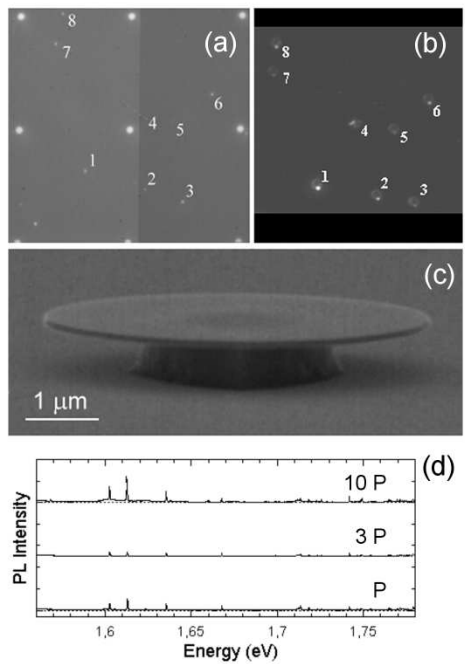}}
\caption{Micro-photoluminescence images of (a) unprocessed sample with gold marks and (b) microdisk sample, with each QD assigned a number. A SEM-image (c) of the microdisk containing QD 1 and spectra (d) from this disk at different low pumping levels. }
\end{figure}

Photoluminescence experiments were done in a liquid helium cooled cryostat at a temperature of 10 K. The luminescence was collected using a microscope objective with 0.4 numerical aperture. The excitation source was a frequency doubled Nd:YAG laser emitting at 532 nm. Spectra were acquired with a cooled CCD and a monochromator. Images were taken with a video camera and a bandpass filter (10 nm FWHM) centered at 720 nm. 

InP quantum dots are especially well suited for this kind of experiment as they have been shown to emit single photons at the ideal wavelength for silicon detectors \cite{ZwillerRed}. Lasing of the GaInP in the disks we study here has been reported \cite{OurLaser}.

Fig. 1d shows spectra obtained on a microdisk containing a single quantum dot. Only a few sharp peaks are visible at low pumping power, which is typical for a single InP QD of this size \cite{PerssonInP}.

The possibility to achieve strong coupling with a single quantum dot has been discussed by G\'erard et al. \cite{StrongCoupling}. Microdisks are the best candidates, as micropillars have too large linewidths and microspheres have too large mode volumes. We predict that by placing a single quantum dot in the maximum field of a microdisk resonator of the right diameter, strong coupling could be achieved for a single quantum dot.

\begin{figure}[h]
\center{\includegraphics[width=1\columnwidth]{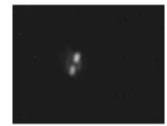}}
\caption{Micro-photoluminescence images of a microdisk containing two QDs close to the edge. }
\end{figure}

In an other microdisk sample, with random distribution of QDs inside the cavities, the structure in Fig. 2 was observed. Such structures would be possible to create in a controlled way by using the demonstrated processing scheme. This might prove particularly useful for evaluating quantum information processing schemes based on this type of structure \cite{Imamoglucomputer}.

\section*{Conclusions}

A technique to position single quantum dots in microdisks has been developed. Optical studies have confirmed the presence of only one QD in the microdisk. One major application of this technique is the possibility to compare theoretical and experimental measurements on the lifetime modifications, as the coupling of the dot to the field depends strongly on the dot position. Further studies should enable us to reach the strong coupling regime for a single dot in a microdisk.

\section*{Acknowledgments}

Part of this work was performed within the Nanometer Structure Consortium in Lund, Sweden and was supported by NFR, TFR, NUTEK and SSF. One of us (VZ) acknowledges funding from the European Union's Marie Curie fellowship.



\end{document}